\documentclass[amsmath,amssymb,showpacs,nofootinbib,twocolumn,prdrc]{revtex4-1}
\usepackage{amsmath,amsfonts}
\usepackage{hyperref}
\usepackage{enumerate}

\begin{document}

\title{Black Brane Entropy and Hydrodynamics} 

\author{Ivan Booth}
\email[]{ibooth@mun.ca}
\affiliation{Department of Mathematics and Statistics, Memorial University of Newfoundland\\
St. John's, Newfoundland and Labrador, A1C 5S7, Canada}

\author{Michal P.~Heller}
\email[]{m.p.heller@uva.nl}
\altaffiliation[On leave from: ]{\it So{\l}tan Institute for Nuclear Studies,
  Ho{\.z}a 69, 00-681 Warsaw, Poland} 
\affiliation{
\it Instituut voor Theoretische Fysica, Universiteit van Amsterdam\\ 
\it Science Park 904, 1090 GL Amsterdam, The Netherlands}

\author{Micha\l\ Spali\'nski}
\email[]{michal.spalinski@fuw.edu.pl}
\affiliation{
\it So{\l}tan Institute for Nuclear Studies, Ho{\.z}a 69, 00-681 Warsaw,
Poland \\ 
\it and Physics Department, University of Bialystok, 15-424 Bialystok,
Poland. } 

\pacs{11.25.Tq, 12.38.Mh}

\newcommand{\ud}{\mathrm{d}}
\newcommand{\bea}{\begin{eqnarray}}
\newcommand{\beal}[1]{\begin{eqnarray}\label{#1}}
\newcommand{\eea}{\end{eqnarray}} 
\newcommand{\be}{\begin{equation}} 
\newcommand{\bel}[1]{\begin{equation}\label{#1}}
\newcommand{\ee}{\end{equation}} 
\newcommand{\rf}[1]{(\ref{#1})}
\newcommand{\nn}{\nonumber}
\newcommand{\bit}{\begin{itemize}}
\newcommand{\eit}{\end{itemize}}
\newcommand{\ben}{\begin{enumerate}}
\newcommand{\een}{\end{enumerate}}
\newcommand{\no}{\noindent}
\newcommand{\req}[1]{(\ref{#1})}
\newcommand{\spacer}{\vspace{6 pt}}

\newcommand{\map}[3]{{#1}:{#2}\rightarrow{#3}}
\newcommand{\Lie}{\mathcal{L}}
\newcommand{\DD}{\mathcal{D}}
\newcommand{\cV}{V}
\newcommand{\tl}{\theta_{(\ell)}}
\newcommand{\tn}{\theta_{(n)}}
\newcommand{\tv}{\theta_{(v)}}
\newcommand{\GG}{{\cal{G}}}
\newcommand{\VV}{{\cal{V}}}

\newcommand{\hnone}{{\cal H}_1}
\newcommand{\hntwo}{{\cal H}_2}
\newcommand{\knone}{{\cal K}_1}
\newcommand{\kntwo}{{\cal K}_2}
\newcommand{\eln}{{\cal L}}
\newcommand{\efn}{{\cal F}}

\newcommand{\xxx}[1]{{\bf #1}}

\def\t{\tilde}
\def\del{\partial}
\def\d{\partial}
\def\half{\frac{1}{2}}
\def\third{\frac{1}{3}}
\def\quart{\frac{1}{4}}

\begin{abstract}

Recent advances in holography have led to the formulation of fluid-gravity
duality, a remarkable connection between the hydrodynamics of certain strongly
coupled media and dynamics of higher dimensional black holes. This paper
introduces a correspondence between phenomenologically defined entropy
currents in relativistic hydrodynamics and ``generalized horizons'' of
near-equilibrium black objects in a dual gravitational description. A general
formula is given, expressing the divergence of the entropy current in terms of
geometric objects which appear naturally in the gravity dual geometry. The
proposed definition is explicitly covariant with respect to boundary
diffeomorphisms and reproduces known results when evaluated for the event
horizon.

\end{abstract}

\maketitle

\section{Introduction}

The holographic representation of field theories
developed following the original formulation of the AdS/CFT correspondence
\cite{Maldacena:1997re} has led to an effective description of strongly
coupled media in, as well as out of, equilibrium. The best understood example
of the latter type is provided by fluid-gravity duality
\cite{Bhattacharyya:2008jc}, which maps the hydrodynamics of strongly coupled
holographic plasmas to the long-wavelength dynamics of black branes appearing
as classical solutions of string or M theory. This description has found
remarkable applications to the physics of QCD plasma experimentally studied at
the Relativistic Heavy Ion Collider and the LHC.

From the hydrodynamics framework comes the notion of an entropy current: a
phenomenological generalization of equilibrium entropy. Such a current is
constructed in the hydrodynamic gradient expansion by requiring that it
reproduces thermodynamic entropy in equilibrium situations and has
non-negative divergence when evaluated on solutions of the hydrodynamic
equations. However, these requirements are not sufficient to specify a unique
current and beyond leading order in gradients there appears to be a family of
currents \cite{Romatschke:2009kr}. 

The aim of this paper is to provide the dual interpretation of ambiguities
affecting the local rate of entropy production within the framework of
fluid-gravity duality. This issue was first raised in
\cite{Bhattacharyya:2008xc}, where a definition of the hydrodynamic entropy
current was given in terms of the area form on the black brane event horizon.
A key element of that proposal is that the divergence of such an entropy
current is guaranteed to be non-negative by the area increase theorem. The
event horizon is an IR concept and in order to associate points on the event
horizon with points at the boundary, a so-called bulk-boundary map is
needed. A choice of the bulk-boundary map clearly will not affect the
non-negativity of the entropy current's divergence, but it can modify the
value of the divergence. The ambiguity in the divergence of the entropy
current constructed in \cite{Bhattacharyya:2008xc} was entirely due to
nonuniqueness of such a bulk-boundary map.

There is however a more fundamental source of ambiguity, which comes from the
fact that, apart from the event horizon, there are other hypersurfaces for
which an area increase theorem holds. For example, in the gravity literature
much attention has been devoted to the study of so-called quasilocal horizons,
since in contrast to the event horizon they may be defined without determining
the full causal structure of spacetime (see
\cite{Ashtekar:2004cn,Booth:2005qc} for reviews). In static situations one of
the 
quasilocal horizons typically coincides with the event horizon. In dynamical
cases these notions differ, but in the near-equilibrium regime relevant to 
fluid-gravity duality they are expected to be close
\cite{Nielsen:2010gm,Booth:2010eu}. It is thus is natural to expect that
quasilocal horizons lead to different hydrodynamic entropy currents in the
boundary field theory
\cite{Bhattacharyya:2008xc,Figueras:2009iu}. Furthermore, other bulk surfaces
may also play a role in the phenomenological construction of the entropy
current in the dual field theory. In particular, timelike surfaces (often
considered in the context of the membrane paradigm) located outside (but close
to) the event horizon may also lead to acceptable entropy currents, as
discussed in \cite{Booth:2009ct} for the case of boost-invariant hydrodynamics
\cite{Janik:2005zt,Heller:2009zz}.

This paper examines the asymptotically AdS gravity dual to second order
conformal fluid dynamics in arbitrary dimension \cite{Bhattacharyya:2008mz}
and demonstrates a correspondence between boundary entropy currents and bulk
hypersurfaces which satisfy generalized area theorems and asymptote to the
event horizon at late time.  Such hypersurfaces (``generalized horizons''),
provide a new source of ambiguity in dual hydrodynamic entropy currents.

\section{Gravitational notions of entropy}

In its simplest form, the bulk
counterpart of a hydrodynamic entropy current should be a gravitational
concept which in equilibrium reduces to the standard event horizon and in
dynamical situations obeys a suitable generalization of the second law of
thermodynamics.  The latter requirement should be equivalent to the
non-negativity of the divergence of a dual entropy current and via the
bulk-boundary map should relate features of the bulk construction to
coefficients appearing in the gradient expansion of a boundary entropy
current. The event horizon provides one concrete realization of this notion
\cite{Bhattacharyya:2008xc}: the equilibrium limit is trivially satisfied and
the generalized second law of thermodynamics is provided by the famous area
theorem stating that the area of spatial sections of the event horizon is
nondecreasing under the evolution along the generators of the horizon.

This example suggests that within the realm of two-derivative gravity, the
minimal structure needed to accommodate the various boundary entropy currents
in a single framework is a hypersurface $\Delta$ along with an associated
tangent vector field $v$: $\Delta$ is a generalization of the event horizon
and $v$ is an analogue of its generators. If this vector field satisfies the
Frobenius condition, then there is a slicing of the horizon into a family of
codimension-two spatial hypersurfaces (``horizon sections'') orthogonal to $v$
with area density $\sqrt{h}$. The generalized second law of thermodynamics
will take the form of the following area increase theorem
\bel{ineq}
\tv = \frac{1}{\sqrt{h}} \mathcal{L}_{v} \sqrt{h} \geq 0 .
\ee
The 
relevant hypersurfaces $\Delta$ 
are those whose equilibrium limit coincides
with the event horizon and for 
which \eqref{ineq} holds by virtue of the dynamics of
general relativity. 

Such generalized area theorems follow from the properties of light rays
propagating in directions normal to the ``horizon sections''.  The normal
space to these sections is spanned by $v$ along with a vector normal to
$\Delta$, denoted here by $m$.  The normalization can be chosen so that $m$ is
exact and $m^2+v^2=0$.  
In the following section it is shown that for each $\Delta$, 
\rf{ineq} is equal to
the divergence of some boundary entropy current.  Thus the gravitational
construction introduced here parallels the phenomenological definition of the
boundary entropy current.

Introducing the ingoing and
outgoing future-pointing null vectors normal to ``horizon sections'' 
reveals the connection with the framework of quasilocal horizons
\cite{Ashtekar:2004cn,Booth:2005qc}.  
Those vectors, denoted respectively by
$\ell$ and $n$ and conventionally normalized so as $\ell \cdot n = -1$, are
given by linear combinations of $v$ and $m$
\be
v = \ell - C n \,\,\,\,\,\, \mathrm{and} \,\,\,\,\,\, m = \ell + C n.
\ee
The scalar $C$ is called the \emph{evolution parameter}.
In terms of $\ell$ and $n$ the inequality \rf{ineq} defining $\Delta$
can be 
equivalently written in a more familiar way in terms of expansions in
directions $\ell$ and $n$
\bel{ineqq}
\frac{1}{\sqrt{h}}\Lie_v \sqrt{h} = \tl - C \tn \geq 0 ,
\ee
where the expansion for a vector field $X$ is defined by 
\bel{thetax}
\theta_{(X)} = \tilde{q}^{ab} \nabla_a X_b =  \mathcal{L}_{X}  \log\sqrt{h} 
\ee
and $\tilde{q}$ is the induced metric on the leaves:
\bel{qdef}
 \tilde{q}_{ab} = g_{ab} + \ell_a n_b + n_a \ell_b \ .
\ee
Two well known choices of $\Delta$ which guarantee \rf{ineq} are the
event horizon, for which $C=0$ and $\tl\geq0$ and a future outer trapping
horizon\footnote{One also requires
  that $\Lie_{n} \tl < 0$, so that there be fully trapped surfaces near
  $\Delta$\cite{Ashtekar:2004cn,Booth:2005qc}.}, for which $C \geq 0$, $\tl = 
0$ and $\tn < 0$. The latter is 
often referred to as an apparent horizon, and this terminology will be used in 
the following.  

The essence of the proposal introduced in this paper is that
\rf{ineq} gives a characterization of bulk hypersurfaces which is
dual to the notion of entropy current provided that the hypersurfaces in
question asymptote to the event horizon. Dynamic event
and apparent horizons might then be regarded as particular realizations of this
idea, though from the perspective of the phenomenological definition, none of
the entropy currents defined this way is distinguished (in line with the
arguments made in \cite{Booth:2009ct}).

\section{General hydrodynamic entropy current and its gravity dual}

Fluid-gravity duality provides a 
setting in which the idea outlined in the previous section is realized. For
simplicity, this paper focuses  on uncharged conformal fluid dynamics in $d$
dimensions and its 
gravity dual. This asymptotically AdS geometry has been explicitly constructed
up to second 
order in gradients in \cite{Bhattacharyya:2008mz}. The form of the metric is 
\bel{wimetric}
\ud s^2 
= -2 u_\mu \ud x^\mu (\ud r + \VV_\alpha \ud
x^\alpha) + \GG_{\mu\nu} \ud x^\mu \ud x^\nu  
\ee
where $r$ is the radial direction in the bulk and $\mu=0,\dots,d-1$. The
components  $G_{ab}$ of the metric \rf{wimetric} depend on boundary directions  
$x^{\mu}$ only via $u^{\mu}$ and $T$, which have the interpretation of local 
velocity and temperature of the dual fluid, as well as the boundary metric
$g_{\mu \nu}$. The geometry is constructed systematically in the gradient
expansion in derivatives of $u^{\mu}$, $T$ and $g_{\mu \nu}$. The leading
order solution to Einstein's equations is a locally boosted and dilated 
(by $b = d/4\pi T$) black brane
\bel{ordzero}
\VV^{\left( 0 \right)}_\mu = \half r^2\big(1 - \frac{1}{\left( r b
  \right)^{d}}\big) u_\mu \quad \mathrm{and} \quad \GG^{\left( 0
  \right)}_{\mu\nu} = r^2 P_{\mu \nu},
\ee
where $P_{\mu \nu}=g_{\mu\nu} + u_\mu u_\nu$ is the projector to the local
rest frame of the fluid. An important simplifying property of the solution
obtained in \cite{Bhattacharyya:2008mz} is the manifest realization of
Weyl covariance \cite{Loganayagam:2008is}, which is a symmetry of conformal
hydrodynamics (up to the Weyl anomaly which appears in even dimensions at order $d$ in
the gradient expansion \cite{Baier:2007ix}). This symmetry
severely restricts the structure and number of allowed gradient terms.

The near-equilibrium nature of the geometry \eqref{wimetric} manifests itself
in the fact that the (future) event horizon of the leading order solution 
$\rf{ordzero}$ is located at $r = 1/b$.
This suggests considering 
more general hypersurfaces $\Delta$ defined as level sets of a function
$S(r,x) = r b(x) + g(x)$, where $g$ is a Weyl-invariant function which can be
written as a linear combination of Weyl-invariant scalars available at each
order of the gradient expansion.

The covector normal to $\Delta$ is 
\bel{mds}
m = dS 
\ee
On $\Delta$ one can introduce coordinates $y^\mu$ ($\mu=0,\dots,d-1$), such
that $S(r(y), x(y)) \equiv const$,  
which implies that tangent vectors to the horizon are of the form 
\bel{tantodelta}
 v =  v^\mu \frac{\partial x^\beta}{\partial y^\mu} \left\{
\frac{\partial}{\partial x^\beta} - 
\left(
\frac{\partial S}{\partial r}
\right)^{-1} 
\left(
\frac{\partial S}{\partial x^\beta}
\right)
 \frac{\partial}{\partial r} 
\right\} \ .
\ee
It is very convenient and physically well motivated to choose the coordinates
on the horizon $y^\mu = x^\mu$.
The position of the horizon can then be expressed in the form
$r=r_\mathrm{H}(x)$. This choice, made already in
\cite{Bhattacharyya:2008xc}, uses bulk causality and connects boundary with
bulk points along ingoing null geodesics emanating from the boundary in the
direction specified locally by $u^{\mu}$ \cite{Bhattacharyya:2008xc}. In the
gauge \eqref{wimetric}, this bulk-boundary map acts trivially relating
points at different radial position and the same $x^{\mu}$ coordinates. It is
clear that beyond leading order in the gradient expansion such a 
bulk-boundary map can be modified in two ways: by changing the direction of
null geodesics emanating from the boundary and by combining a given 
bulk-boundary map with boundary diffeomorphisms
\cite{Bhattacharyya:2008xc}. Up to second order in gradients only the
latter freedom is present and will modify the consequences of the choice
$y^\mu = x^\mu$ precisely in the same way as discussed in 
\cite{Bhattacharyya:2008xc}. 

Vectors tangent to the horizon have a specific $r$ component \rf{tantodelta},
which ensures   
that $m\cdot v = 0$. Thus, choosing the $\mu$ components of the tangent vector 
$v$ determines it fully. These components are constructed as a linear
combination of all the available Weyl-invariant vectors made out of the
hydrodynamic observables and boundary curvature.
The coefficients appearing in 
this gradient expansion of
$v$ are constrained by the imposed normalization
\bel{vnorm}
m^2 + v^2 = 0
\ee
and the
requirement of hypersurface orthogonality. Up to second order
\cite{Booth:2011qy} these coefficients are completely
fixed and so the only quantities defining
the gravitational construction are those appearing in $r_\mathrm{H}(x)$. 
This is
consistent with what is known about two important special cases: if $\Delta$
is the event horizon, then $v$ is tangent to its generators and in this sense
is determined (up to normalization) by $r_{\mathrm{H}}(x) =
r_{\mathrm{EH}}(x)$; if $\Delta$ is 
a spacelike apparent horizon\footnote{This should properly
  be called a future outer trapping horizon.} there is 
a unique foliation\cite{Ashtekar:2005ez} 
and so $v$ is 
again specified up to a normalization by $r_{\mathrm{H}}(x) =
r_{\mathrm{AH}}(x)$. 

The key fact which leads to the general formula for the hydrodynamic entropy
current is that the expansion $\tv$ is simply the divergence
of $v$: 
\bel{vdiv}
\tv = \nabla_a v^a \ .
\ee
To show this, note that the definition of $\tv$ (in the form \rf{thetax}) can
be rewritten in terms of $m$ and $v$ as
\bel{tvv}
\tv = \nabla_a v^a + \frac{1}{C} \left(m^a m^b - v^a v^b\right) \nabla_a v_b
\ .
\ee
The second term above can be shown to vanish by applying two identities which
follow by differentiating the normalization condition \rf{vnorm}
and the fact that $m$ is exact \rf{mds}.  Since $v$ is tangent to the horizon,
its divergence in the adopted coordinate   
system can be written in a way which makes it trivial to map the relation
\rf{vdiv} to the boundary. To this end, note that \rf{vdiv} can be expressed 
in the form
\bel{tvdel}
\tv = \frac{1}{\sqrt{-G}} \frac{\partial}{\partial x^a} \left(\sqrt{-G}
v^a\right) \ ,
\ee
where $G$ is the determinant of the bulk metric. 
This relation is to be evaluated on $\Delta$. The derivatives with respect
to the bulk coordinates $x^\mu$ can be expressed in terms of derivatives with
respect to the coordinates  $y^\mu$ intrinsic to $\Delta$, and $v_r$ can be
eliminated using \rf{tantodelta}, which leads to
\be
\tv = b \frac{1}{\sqrt{-G}} \frac{\partial}{\partial y^\mu} \left(
 \frac{1}{b} \sqrt{-G} v^\mu \right) \ .
\ee  
Finally, this can be expressed in terms of the Levi-Civita connection
associated with the boundary metric $g$ as
\bel{tvbdry}
\tv = b \sqrt{\frac{g}{G}} \nabla_\mu \left( \frac{1}{b} \sqrt{\frac{G}{g}} \ 
v^\mu \right) \ .
\ee  
Because of the bulk-boundary map introduced earlier, \rf{tvbdry} can be
interpreted as an equation pertaining to 
the boundary field theory. This motivates the definition of the hydrodynamic
entropy current as
\bel{entrocur}
J^\mu = \frac{1}{4 G_{N}} \frac{1}{b} \sqrt{\frac{G}{g}}\ v^\mu \ ,
\ee
where $G_{N}$ is a
$(d+1)$-dimensional Newton's constant introduced to reproduce thermodynamic
entropy in the leading order and AdS radius is set to 1, as in
\eqref{ordzero}. 
The point of this definition is that it follows from \rf{tvbdry} that 
\bel{ecdiv}
\nabla_\mu J^\mu = \frac{1}{4 G_{N}} \frac{1}{b} \sqrt{\frac{G}{g}}\ \tv \ . 
\ee
Thus the divergence of the current \rf{entrocur} is indeed 
non-negative if and only if the hypersurface $\Delta$ satisfies $\rf{ineq}$.  
Note that one can always shift the entropy current \rf{entrocur} by a
vector whose divergence vanishes without changing \rf{ecdiv}. This possibility
follows from the freedom to shift 
\bel{vshift}
v \rightarrow v + w
\ee
in \rf{vdiv}, where $w$ is tangent to $\Delta$ and $\nabla_a w^a = 0$.

Formula \rf{entrocur} [together with \rf{ineq}] is the main
result reported in this paper. It provides a map between hypersurfaces 
in the IR 
region of the geometry of fluid-gravity duality and phenomenologically defined 
entropy currents. 
It is explicitly covariant under boundary
diffeomorphisms and applies to both apparent and event horizons, as well
as other (timelike or spacelike) hypersurfaces, provided they satisfy
$\tv \geq 0$ (or equivalently $\nabla_{\mu} J^{\mu} \geq 0$).

\section{Entropy currents and dual hypersurfaces at second order}

The geometry of fluid-gravity duality has been explicitly constructed 
only up
to second order in gradients. To that order the defining condition for
the hypersurface $\Delta$ contains three parameters multiplying the three
independent hydrodynamic scalars 
\bel{rh2ord}
r_{\mathrm{H}} = \frac{1}{b} + b\left(h_{1} \sigma_{\mu \nu} \sigma^{\mu \nu} + 
h_{2} \omega_{\mu \nu} \omega^{\mu \nu} + h_{3} {\cal R} \right), 
\ee
where $\sigma$, $\omega$ and ${\cal R}$ are the shear tensor, vorticity and
boundary Ricci scalar, as defined in \cite{Bhattacharyya:2008mz}. 
The vector $v$ is completely fixed \cite{Booth:2011qy} by the
self-consistency of the bulk construction discussed earlier; up to second
order  
\bel{v2ord} 
v^{\mu} = b u^{\mu} +  \frac{1}{d-2} b^{3} P^{\mu\nu} \left(
 \frac{2}{d} \DD^\rho \sigma_{\nu \rho} + 
\DD^\rho \omega_{\nu \rho}
\right) 
\ee
where $\DD$ is the Weyl-covariant derivative (see \cite{Bhattacharyya:2008mz}
for details). The $v^{r}$ component is fixed by requiring that $v$ is tangent
to $\Delta$ [see \rf{tantodelta}].

The current \rf{entrocur} dual to $r_{\mathrm{H}}(x)$ given in \eqref{rh2ord} reads
\beal{ecres}
J^\mu &=&  \frac{1}{4 G_{N}} b^{1-d} \left( u^\mu \right. + \\ \nonumber  
&+& \left. b^2 P^{\mu\nu} (j^\perp_1 \DD^\rho  
\sigma_{\nu \rho} + j^\perp_2 \DD^\rho \omega_{\nu \rho}) + \right. \\ \nonumber
&+& \left. b^2 (j^{\parallel}_1 \sigma_{\alpha \beta} \sigma^{\alpha \beta}  +
j^{\parallel}_2 \omega_{\alpha \beta} \omega^{\alpha \beta}  + 
j^{\parallel}_3 {\cal R}) u^\mu \right) ,
\eea
where $j^{\perp}_1$ and $j^{\perp}_2$ are fixed and equal
\bel{jperp}
j^\perp_1 = \frac{2}{d(d-2)},  \quad j^\perp_2 =  \frac{1}{d-2} \ ,
\ee
while the $j^{\parallel}_{i}$'s depend on the choice of $h_{i}$'s
\beal{jII}
j^{\parallel}_{1} &=& (d-1) h_{1} - K_1(1), \nonumber \\ 
j^{\parallel}_{2} &=& (d-1) h_2 + \half \quad \mathrm{and} 
\quad j^{\parallel}_{3} = (d-1) h_{3} .
\eea
Here $K_1$ [as well as $K_2$ in eqn. \rf{ecehcoefs} below] is a function
defined in 
\cite{Bhattacharyya:2008mz}. 
The current obeys a generalized second law of
thermodynamics if and only if the dual surface $\Delta$ satisfies
\eqref{ineq}.

To get a feeling for what this entails,  it is interesting to compare the
entropy currents which follow from taking  
the $h_{i}$'s computed for the event and 
apparent horizon \cite{Booth:2011qy}.
For the event horizon one finds 
\beal{ecehcoefs}
j^{\parallel}_1 &=& - K_1(1)  - \frac{1}{d} K_2(1) + \frac{2(d-1)}{d (d-2)}
\nonumber \\  
j^{\parallel}_2 &=&  \frac{(2-3 d)}{2 d (d-2)} \quad \mathrm{and} \quad 
j^{\parallel}_3 = - \frac{1}{d (d-2)}, 
\eea
which matches the result of \cite{Bhattacharyya:2008mz}. One can similarly
evaluate the entropy current on the apparent horizon. One again finds
\rf{ecres} with the coefficients given by \eqref{jperp} and \eqref{ecehcoefs},
apart from $j^{\parallel}_1$, which differs from its value on the event
horizon by $-\frac{4}{d^2}$. Note that $j^{\parallel}_{1}$ is the only
unconstrained parameter in the analysis presented in \cite{Romatschke:2009kr}
which appears in the divergence of the current.  Changing $j^{\parallel}_{1}$
coincides with modifying the bulk-boundary map given by $y^{\mu} = x^{\mu}$ in
the gauge \eqref{wimetric} by including boundary diffeomorphisms
\cite{Bhattacharyya:2008xc}. It is natural to speculate that changing
$j^{\parallel}_{1}$ continuously within a given bulk-boundary map might be
related to a one-parameter family of second order entropy currents. These
would correspond to a set of timelike, spacelike or null bulk surfaces
satisfying area theorem at the level of the gradient expansion, in line with
the proposal of \cite{Booth:2009ct}.

Note finally, that the freedom \rf{vshift} (which preserves the divergence of
the entropy current) translates to a shift of $j^{\parallel}_2, j^\perp_2$
with $\delta j^{\parallel}_2 + \delta j^\perp_2 = 0$. This is the same as the
ambiguity noted in\cite{Bhattacharyya:2008xc}.

\section{Conclusions}

This paper proposes a direct link between
phenomenologically defined entropy currents in holographic field theories and
generalized horizons in the dual gravity background within the
near-equilibrium regime. These geometrical objects
are defined by the area theorem and the requirement of asymptoting to the
event horizon. Such a definition matches precisely the defining condition of
the hydrodynamic entropy current and embraces in a common framework both the
event and 
apparent horizons within fluid-gravity duality. The
geometric formula for the dual entropy current is manifestly covariant in the
boundary sense and in the case of the event horizon leads to results which
agree with \cite{Bhattacharyya:2008xc} and \cite{Bhattacharyya:2008mz}.

It is further stipulated that at second order in the gradient expansion there
is a one-parameter family of entropy currents arising in this way. This
freedom arises solely from the freedom of choosing different
$r_{\mathrm{H}}(x)$ to define a dual entropy current and needs to be
supplemented with the known freedom of defining a bulk-boundary map, as well
as adding terms with vanishing divergence.  Apart from these, the surface
itself is the only geometric object relevant for the dual entropy current: it
is conjectured (and verified to second order of the gradient expansion) that
the foliation of the horizon is uniquely fixed within the gradient
expansion. This conclusion is consistent with the results of the analysis
presented in \cite{Romatschke:2009kr}.

From the point of view of the phenomenological definition, none of the entropy
currents introduced here is distinguished. Perhaps extending the requirements
(by e.g. causality) will rule out some of them. Note in particular that the
entropy current defined by the apparent horizon in a given bulk-boundary map
is equivalent (up to second order in the gradient expansion) to the entropy
current defined by the event horizon in another bulk-boundary map. Since the
latter is known to be acausal, it is clear that there should be an interplay
between causality of dual entropy currents and the choice of bulk-boundary
map. It would be fascinating to investigate this issue further, especially
since the apparent horizon in the fluid-gravity duality seems to be free of
foliation-dependence.
It would also be very interesting to generalize the
construction presented in this paper to cases of gravity duals to
charged\cite{Banerjee:2008th,Erdmenger:2008rm} and 
nonconformal fluid dynamics\cite{Kanitscheider:2009as}.

\vspace{0.7cm}

\centerline{\textbf{Acknowledgments}}

\vspace{0.5cm}

The authors acknowledge the use of Kasper Peeters's
excellent package {\tt Cadabra}
\cite{DBLP:journals/corr/abs-cs-0608005,Peeters:2007wn}. This work was
partially supported by Polish Ministry of Science and Higher Education grants
\emph{N N202 173539} and \emph{N N202 105136}.  MH acknowledges support from
FNP and FOM. IB was supported by NSERC.  The authors would like to thank
R. A. Janik, S. Minwalla and M. Rangamani for helpful comments on the
manuscript.

\bibliographystyle{apsrev4-1}
\bibliography{biblio.bib}

%Merlin.mbs v4.21 2009-07-09.
\begin{thebibliography}{10}%
\makeatletter
\providecommand \@ifxundefined [1]{%
 \ifx #1\undefined \expandafter \@firstoftwo
 \else \expandafter \@secondoftwo
\fi
}%
\providecommand \@ifnum [1]{%
 \ifnum #1\expandafter \@firstoftwo
 \else \expandafter \@secondoftwo
\fi
}%
\providecommand \enquote [1]{``#1''}%
\providecommand \bibnamefont  [1]{#1}%
\providecommand \bibfnamefont [1]{#1}%
\providecommand \citenamefont [1]{#1}%
\providecommand\href[0]{\@sanitize\@href}%
\providecommand\@href[1]{\endgroup\@@startlink{#1}\endgroup\@@href}%
\providecommand\@@href[1]{#1\@@endlink}%
\providecommand \@sanitize [0]{\begingroup\catcode`\&12\catcode`\#12\relax}%
\@ifxundefined \pdfoutput {\@firstoftwo}{%
 \@ifnum{\z@=\pdfoutput}{\@firstoftwo}{\@secondoftwo}%
}{%
 \providecommand\@@startlink[1]{\leavevmode\special{html:<a href="#1">}}%
 \providecommand\@@endlink[0]{\special{html:</a>}}%
}{%
 \providecommand\@@startlink[1]{%
  \leavevmode
  \pdfstartlink
   attr{/Border[0 0 1 ]/H/I/C[0 1 1]}%
   user{/Subtype/Link/A<</Type/Action/S/URI/URI(#1)>>}%
  \relax
 }%
 \providecommand\@@endlink[0]{\pdfendlink}%
}%
\providecommand \url  [0]{\begingroup\@sanitize \@url }%
\providecommand \@url [1]{\endgroup\@href {#1}{\urlprefix}}%
\providecommand \urlprefix [0]{URL }%
\providecommand \Eprint[0]{\href }%
\@ifxundefined \urlstyle {%
  \providecommand \doi [1]{doi:\discretionary{}{}{}#1}%
}{%
  \providecommand \doi [0]{doi:\discretionary{}{}{}\begingroup
  \urlstyle{rm}\Url }%
}%
\providecommand \doibase [0]{http://dx.doi.org/}%
\providecommand \Doi[1]{\href{\doibase#1}}%
\providecommand \bibAnnote [3]{%
  \BibitemShut{#1}%
  \begin{quotation}\noindent
    \textsc{Key:}\ #2\\\textsc{Annotation:}\ #3%
  \end{quotation}%
}%
\providecommand \bibAnnoteFile [2]{%
  \IfFileExists{#2}{\bibAnnote {#1} {#2} {\input{#2}}}{}%
}%
\providecommand \typeout [0]{\immediate \write \m@ne }%
\providecommand \selectlanguage [0]{\@gobble}%
\providecommand \bibinfo [0]{\@secondoftwo}%
\providecommand \bibfield [0]{\@secondoftwo}%
\providecommand \translation [1]{[#1]}%
\providecommand \BibitemOpen[0]{}%
\providecommand \bibitemStop [0]{}%
\providecommand \bibitemNoStop [0]{.\EOS\space}%
\providecommand \EOS [0]{\spacefactor3000\relax}%
\providecommand \BibitemShut [1]{\csname bibitem#1\endcsname}%
%</preamble>
\bibitem{Maldacena:1997re}%
  \BibitemOpen
  \bibfield{author}{%
  \bibinfo {author} {\bibfnamefont{J.~M.}\ \bibnamefont{Maldacena}},\ }%
  \bibfield{journal}{%
  \bibinfo {journal} {Adv. Theor. Math. Phys.}\ }%
  \textbf{\bibinfo {volume} {2}},\ \bibinfo {pages} {231} (\bibinfo {year}
  {1998}),\ \Eprint{http://arxiv.org/abs/hep-th/9711200}{arXiv:hep-th/9711200}%
  \bibAnnoteFile{NoStop}{Maldacena:1997re}%
%%CITATION = HEP-TH/9711200;%%
\bibitem{Bhattacharyya:2008jc}%
  \BibitemOpen
  \bibfield{author}{%
  \bibinfo {author} {\bibfnamefont{S.}~\bibnamefont{Bhattacharyya}}, \bibinfo
  {author} {\bibfnamefont{V.~E.}\ \bibnamefont{Hubeny}}, \bibinfo {author}
  {\bibfnamefont{S.}~\bibnamefont{Minwalla}},\ and\ \bibinfo {author}
  {\bibfnamefont{M.}~\bibnamefont{Rangamani}},\ }%
  \bibfield{journal}{%
  \Doi{10.1088/1126-6708/2008/02/045}{\bibinfo {journal} {JHEP}}\ }%
  \textbf{\bibinfo {volume} {02}},\ \bibinfo {pages} {045} (\bibinfo {year}
  {2008}),\ \Eprint{http://arxiv.org/abs/0712.2456}{arXiv:0712.2456 [hep-th]}%
  \bibAnnoteFile{NoStop}{Bhattacharyya:2008jc}%
%%CITATION = 0712.2456;%%
\bibitem{Romatschke:2009kr}%
  \BibitemOpen
  \bibfield{author}{%
  \bibinfo {author} {\bibfnamefont{P.}~\bibnamefont{Romatschke}},\ }%
  \bibfield{journal}{%
  \Doi{10.1088/0264-9381/27/2/025006}{\bibinfo {journal} {Class. Quant.
  Grav.}}\ }%
  \textbf{\bibinfo {volume} {27}},\ \bibinfo {pages} {025006} (\bibinfo {year}
  {2010}),\ \Eprint{http://arxiv.org/abs/0906.4787}{arXiv:0906.4787 [hep-th]}%
  \bibAnnoteFile{NoStop}{Romatschke:2009kr}%
%%CITATION = 0906.4787;%%
\bibitem{Bhattacharyya:2008xc}%
  \BibitemOpen
  \bibfield{author}{%
  \bibinfo {author} {\bibfnamefont{S.}~\bibnamefont{Bhattacharyya}}
  \emph{et~al.},\ }%
  \bibfield{journal}{%
  \bibinfo {journal} {JHEP}\ }%
  \textbf{\bibinfo {volume} {06}},\ \bibinfo {pages} {055} (\bibinfo {year}
  {2008}),\ \Eprint{http://arxiv.org/abs/0803.2526}{arXiv:0803.2526 [hep-th]}%
  \bibAnnoteFile{NoStop}{Bhattacharyya:2008xc}%
%%CITATION = 0803.2526;%%
\bibitem{Ashtekar:2004cn}%
  \BibitemOpen
  \bibfield{author}{%
  \bibinfo {author} {\bibfnamefont{A.}~\bibnamefont{Ashtekar}}\ and\ \bibinfo
  {author} {\bibfnamefont{B.}~\bibnamefont{Krishnan}},\ }%
  \bibfield{journal}{%
  \bibinfo {journal} {Living Rev. Rel.}\ }%
  \textbf{\bibinfo {volume} {7}},\ \bibinfo {pages} {10} (\bibinfo {year}
  {2004}),\ \Eprint{http://arxiv.org/abs/gr-qc/0407042}{arXiv:gr-qc/0407042}%
  \bibAnnoteFile{NoStop}{Ashtekar:2004cn}%
%%CITATION = GR-QC/0407042;%%
\bibitem{Booth:2005qc}%
  \BibitemOpen
  \bibfield{author}{%
  \bibinfo {author} {\bibfnamefont{I.}~\bibnamefont{Booth}},\ }%
  \bibfield{journal}{%
  \Doi{10.1139/p05-063}{\bibinfo {journal} {Can. J. Phys.}}\ }%
  \textbf{\bibinfo {volume} {83}},\ \bibinfo {pages} {1073} (\bibinfo {year}
  {2005}),\ \Eprint{http://arxiv.org/abs/gr-qc/0508107}{arXiv:gr-qc/0508107}%
  \bibAnnoteFile{NoStop}{Booth:2005qc}%
%%CITATION = GR-QC/0508107;%%
\bibitem{Nielsen:2010gm}%
  \BibitemOpen
  \bibfield{author}{%
  \bibinfo {author} {\bibfnamefont{A.~B.}\ \bibnamefont{Nielsen}},\ }%
  \bibfield{journal}{%
  \Doi{10.1088/0264-9381/27/24/245016}{\bibinfo {journal} {Class. Quant.
  Grav.}}\ }%
  \textbf{\bibinfo {volume} {27}},\ \bibinfo {pages} {245016} (\bibinfo {year}
  {2010}),\ \Eprint{http://arxiv.org/abs/1006.2448}{arXiv:1006.2448 [gr-qc]}%
  \bibAnnoteFile{NoStop}{Nielsen:2010gm}%
%%CITATION = 1006.2448;%%
\bibitem{Booth:2010eu}%
  \BibitemOpen
  \bibfield{author}{%
  \bibinfo {author} {\bibfnamefont{I.}~\bibnamefont{Booth}}\ and\ \bibinfo
  {author} {\bibfnamefont{J.}~\bibnamefont{Martin}},\ }%
  \bibfield{journal}{%
  \Doi{10.1103/PhysRevD.82.124046}{\bibinfo {journal} {Phys. Rev.}}\ }%
  \textbf{\bibinfo {volume} {D82}},\ \bibinfo {pages} {124046} (\bibinfo {year}
  {2010}),\ \Eprint{http://arxiv.org/abs/1007.1642}{arXiv:1007.1642 [gr-qc]}%
  \bibAnnoteFile{NoStop}{Booth:2010eu}%
%%CITATION = 1007.1642;%%
\bibitem{Figueras:2009iu}%
  \BibitemOpen
  \bibfield{author}{%
  \bibinfo {author} {\bibfnamefont{P.}~\bibnamefont{Figueras}}, \bibinfo
  {author} {\bibfnamefont{V.~E.}\ \bibnamefont{Hubeny}}, \bibinfo {author}
  {\bibfnamefont{M.}~\bibnamefont{Rangamani}},\ and\ \bibinfo {author}
  {\bibfnamefont{S.~F.}\ \bibnamefont{Ross}},\ }%
  \bibfield{journal}{%
  \Doi{10.1088/1126-6708/2009/04/137}{\bibinfo {journal} {JHEP}}\ }%
  \textbf{\bibinfo {volume} {04}},\ \bibinfo {pages} {137} (\bibinfo {year}
  {2009}),\ \Eprint{http://arxiv.org/abs/0902.4696}{arXiv:0902.4696 [hep-th]}%
  \bibAnnoteFile{NoStop}{Figueras:2009iu}%
%%CITATION = 0902.4696;%%
\bibitem{Booth:2009ct}%
  \BibitemOpen
  \bibfield{author}{%
  \bibinfo {author} {\bibfnamefont{I.}~\bibnamefont{Booth}}, \bibinfo {author}
  {\bibfnamefont{M.~P.}\ \bibnamefont{Heller}},\ and\ \bibinfo {author}
  {\bibfnamefont{M.}~\bibnamefont{Spalinski}},\ }%
  \bibfield{journal}{%
  \Doi{10.1103/PhysRevD.80.126013}{\bibinfo {journal} {Phys. Rev.}}\ }%
  \textbf{\bibinfo {volume} {D80}},\ \bibinfo {pages} {126013} (\bibinfo {year}
  {2009}),\ \Eprint{http://arxiv.org/abs/0910.0748}{arXiv:0910.0748 [hep-th]}%
  \bibAnnoteFile{NoStop}{Booth:2009ct}%
%%CITATION = 0910.0748;%%
\bibitem{Janik:2005zt}%
  \BibitemOpen
  \bibfield{author}{%
  \bibinfo {author} {\bibfnamefont{R.~A.}\ \bibnamefont{Janik}}\ and\ \bibinfo
  {author} {\bibfnamefont{R.~B.}\ \bibnamefont{Peschanski}},\ }%
  \bibfield{journal}{%
  \Doi{10.1103/PhysRevD.73.045013}{\bibinfo {journal} {Phys. Rev.}}\ }%
  \textbf{\bibinfo {volume} {D73}},\ \bibinfo {pages} {045013} (\bibinfo {year}
  {2006}),\ \Eprint{http://arxiv.org/abs/hep-th/0512162}{arXiv:hep-th/0512162}%
  \bibAnnoteFile{NoStop}{Janik:2005zt}%
%%CITATION = HEP-TH/0512162;%%
\bibitem{Heller:2009zz}%
  \BibitemOpen
  \bibfield{author}{%
  \bibinfo {author} {\bibfnamefont{M.~P.}\ \bibnamefont{Heller}}, \bibinfo
  {author} {\bibfnamefont{P.}~\bibnamefont{Surowka}}, \bibinfo {author}
  {\bibfnamefont{R.}~\bibnamefont{Loganayagam}}, \bibinfo {author}
  {\bibfnamefont{M.}~\bibnamefont{Spalinski}},\ and\ \bibinfo {author}
  {\bibfnamefont{S.~E.}\ \bibnamefont{Vazquez}},\ }%
  \bibfield{journal}{%
  \Doi{10.1103/PhysRevLett.102.041601}{\bibinfo {journal} {Phys. Rev. Lett.}}\
  }%
  \textbf{\bibinfo {volume} {102}},\ \bibinfo {pages} {041601} (\bibinfo {year}
  {2009}),\ \Eprint{http://arxiv.org/abs/0805.3774}{arXiv:0805.3774 [hep-th]}%
  \bibAnnoteFile{NoStop}{Heller:2009zz}%
%%CITATION = PRLTA,102,041601;%%
\bibitem{Bhattacharyya:2008mz}%
  \BibitemOpen
  \bibfield{author}{%
  \bibinfo {author} {\bibfnamefont{S.}~\bibnamefont{Bhattacharyya}}, \bibinfo
  {author} {\bibfnamefont{R.}~\bibnamefont{Loganayagam}}, \bibinfo {author}
  {\bibfnamefont{I.}~\bibnamefont{Mandal}}, \bibinfo {author}
  {\bibfnamefont{S.}~\bibnamefont{Minwalla}},\ and\ \bibinfo {author}
  {\bibfnamefont{A.}~\bibnamefont{Sharma}},\ }%
  \bibfield{journal}{%
  \Doi{10.1088/1126-6708/2008/12/116}{\bibinfo {journal} {JHEP}}\ }%
  \textbf{\bibinfo {volume} {12}},\ \bibinfo {pages} {116} (\bibinfo {year}
  {2008}),\ \Eprint{http://arxiv.org/abs/0809.4272}{arXiv:0809.4272 [hep-th]}%
  \bibAnnoteFile{NoStop}{Bhattacharyya:2008mz}%
%%CITATION = 0809.4272;%%
\bibitem{Loganayagam:2008is}%
  \BibitemOpen
  \bibfield{author}{%
  \bibinfo {author} {\bibfnamefont{R.}~\bibnamefont{Loganayagam}},\ }%
  \bibfield{journal}{%
  \Doi{10.1088/1126-6708/2008/05/087}{\bibinfo {journal} {JHEP}}\ }%
  \textbf{\bibinfo {volume} {05}},\ \bibinfo {pages} {087} (\bibinfo {year}
  {2008}),\ \Eprint{http://arxiv.org/abs/0801.3701}{arXiv:0801.3701 [hep-th]}%
  \bibAnnoteFile{NoStop}{Loganayagam:2008is}%
%%CITATION = 0801.3701;%%
\bibitem{Baier:2007ix}%
  \BibitemOpen
  \bibfield{author}{%
  \bibinfo {author} {\bibfnamefont{R.}~\bibnamefont{Baier}}, \bibinfo {author}
  {\bibfnamefont{P.}~\bibnamefont{Romatschke}}, \bibinfo {author}
  {\bibfnamefont{D.~T.}\ \bibnamefont{Son}}, \bibinfo {author}
  {\bibfnamefont{A.~O.}\ \bibnamefont{Starinets}},\ and\ \bibinfo {author}
  {\bibfnamefont{M.~A.}\ \bibnamefont{Stephanov}},\ }%
  \bibfield{journal}{%
  \Doi{10.1088/1126-6708/2008/04/100}{\bibinfo {journal} {JHEP}}\ }%
  \textbf{\bibinfo {volume} {04}},\ \bibinfo {pages} {100} (\bibinfo {year}
  {2008}),\ \Eprint{http://arxiv.org/abs/0712.2451}{arXiv:0712.2451 [hep-th]}%
  \bibAnnoteFile{NoStop}{Baier:2007ix}%
%%CITATION = 0712.2451;%%
\bibitem{Booth:2011qy}%
  \BibitemOpen
  \bibfield{author}{%
  \bibinfo {author} {\bibfnamefont{I.}~\bibnamefont{Booth}}, \bibinfo {author}
  {\bibfnamefont{M.~P.}\ \bibnamefont{Heller}}, \bibinfo {author}
  {\bibfnamefont{G.}~\bibnamefont{Plewa}},\ and\ \bibinfo {author}
  {\bibfnamefont{M.}~\bibnamefont{Spalinski}}}%
   (\bibinfo {year} {2011}),\
  \Eprint{http://arxiv.org/abs/1102.2885}{arXiv:1102.2885 [hep-th]}%
  \bibAnnoteFile{NoStop}{Booth:2011qy}%
%%CITATION = 1102.2885;%%
\bibitem{Ashtekar:2005ez}%
  \BibitemOpen
  \bibfield{author}{%
  \bibinfo {author} {\bibfnamefont{A.}~\bibnamefont{Ashtekar}}\ and\ \bibinfo
  {author} {\bibfnamefont{G.~J.}\ \bibnamefont{Galloway}},\ }%
  \bibfield{journal}{%
  \bibinfo {journal} {Adv. Theor. Math. Phys.}\ }%
  \textbf{\bibinfo {volume} {9}},\ \bibinfo {pages} {1} (\bibinfo {year}
  {2005}),\ \Eprint{http://arxiv.org/abs/gr-qc/0503109}{arXiv:gr-qc/0503109}%
  \bibAnnoteFile{NoStop}{Ashtekar:2005ez}%
%%CITATION = GR-QC/0503109;%%
\bibitem{Banerjee:2008th}%
  \BibitemOpen
  \bibfield{author}{%
  \bibinfo {author} {\bibfnamefont{N.}~\bibnamefont{Banerjee}}, \bibinfo
  {author} {\bibfnamefont{J.}~\bibnamefont{Bhattacharya}}, \bibinfo {author}
  {\bibfnamefont{S.}~\bibnamefont{Bhattacharyya}}, \bibinfo {author}
  {\bibfnamefont{S.}~\bibnamefont{Dutta}}, \bibinfo {author}
  {\bibfnamefont{R.}~\bibnamefont{Loganayagam}}, \emph{et~al.},\ }%
  \bibfield{journal}{%
  \Doi{10.1007/JHEP01(2011)094}{\bibinfo {journal} {JHEP}}\ }%
  \textbf{\bibinfo {volume} {1101}},\ \bibinfo {pages} {094} (\bibinfo {year}
  {2011}),\ \Eprint{http://arxiv.org/abs/0809.2596}{arXiv:0809.2596 [hep-th]}%
  \bibAnnoteFile{NoStop}{Banerjee:2008th}%
\bibitem{Erdmenger:2008rm}%
  \BibitemOpen
  \bibfield{author}{%
  \bibinfo {author} {\bibfnamefont{J.}~\bibnamefont{Erdmenger}}, \bibinfo
  {author} {\bibfnamefont{M.}~\bibnamefont{Haack}}, \bibinfo {author}
  {\bibfnamefont{M.}~\bibnamefont{Kaminski}},\ and\ \bibinfo {author}
  {\bibfnamefont{A.}~\bibnamefont{Yarom}},\ }%
  \bibfield{journal}{%
  \Doi{10.1088/1126-6708/2009/01/055}{\bibinfo {journal} {JHEP}}\ }%
  \textbf{\bibinfo {volume} {0901}},\ \bibinfo {pages} {055} (\bibinfo {year}
  {2009}),\ \Eprint{http://arxiv.org/abs/0809.2488}{arXiv:0809.2488 [hep-th]}%
  \bibAnnoteFile{NoStop}{Erdmenger:2008rm}%
\bibitem{Kanitscheider:2009as}%
  \BibitemOpen
  \bibfield{author}{%
  \bibinfo {author} {\bibfnamefont{I.}~\bibnamefont{Kanitscheider}}\ and\
  \bibinfo {author} {\bibfnamefont{K.}~\bibnamefont{Skenderis}},\ }%
  \bibfield{journal}{%
  \Doi{10.1088/1126-6708/2009/04/062}{\bibinfo {journal} {JHEP}}\ }%
  \textbf{\bibinfo {volume} {04}},\ \bibinfo {pages} {062} (\bibinfo {year}
  {2009}),\ \Eprint{http://arxiv.org/abs/0901.1487}{arXiv:0901.1487 [hep-th]}%
  \bibAnnoteFile{NoStop}{Kanitscheider:2009as}%
%%CITATION = 0901.1487;%%
\bibitem{DBLP:journals/corr/abs-cs-0608005}%
  \BibitemOpen
  \bibfield{author}{%
  \bibinfo {author} {\bibfnamefont{K.}~\bibnamefont{Peeters}},\ }%
  \bibfield{journal}{%
  \bibinfo {journal} {CoRR}\ }%
  \textbf{\bibinfo {volume} {abs/cs/0608005}} (\bibinfo {year} {2006})%
  \bibAnnoteFile{NoStop}{DBLP:journals/corr/abs-cs-0608005}%
\bibitem{Peeters:2007wn}%
  \BibitemOpen
  \bibfield{author}{%
  \bibinfo {author} {\bibfnamefont{K.}~\bibnamefont{Peeters}}}%
   (\bibinfo {year} {2007}),\
  \Eprint{http://arxiv.org/abs/hep-th/0701238}{arXiv:hep-th/0701238}%
  \bibAnnoteFile{NoStop}{Peeters:2007wn}%
%%CITATION = HEP-TH/0701238;%%
\end{thebibliography}%

\end{document}